\documentclass{PoS}

\usepackage{amssymb}
\usepackage{amsmath}
\usepackage{fontenc}
\usepackage{times}
\usepackage{mathptmx}
\usepackage{graphicx}
\usepackage{lineno}
\usepackage{multirow}

\newcommand{\snn}{\ensuremath{\sqrt{s_{_{\rm NN}}}}}
\newcommand{\pt}{\ensuremath{p_{\rm T}}}
\newcommand{\dd}{\mbox{d}}

\title{Measurement of electroweak-boson production in Pb--Pb and p--Pb collisions at the LHC with ALICE}
\ShortTitle{Electroweak-boson in Pb--Pb and p--Pb collisions with ALICE}
\author{\speaker{G. Taillepied} for the ALICE collaboration\\
  Laboratoire de Physique de Clermont, France\\
  E-mail:  \email{g.taillepied@cern.ch}
}
\FullConference{EPS-HEP 2019\\
  Ghent, Belgium\\
  July, 10--17 2019
}
\abstract{Measurements of electroweak-bosons in heavy-ion collisions with ALICE are reported. They are measured at large rapidities, via their (di)muonic decay: Z $\rightarrow \mu^+ \mu^-$ and W$^\pm \rightarrow \mu^\pm \nu$. The Z production and nuclear modification factor in Pb--Pb collisions at \snn{} = 5.02 TeV are presented and compared to theoretical predictions with or without nuclear modifications of the Parton Distribution Functions. Measurements of electroweak-boson production in p--Pb collisions at \snn{} = 5.02 and 8.16 TeV are also presented and discussed.}

\begin{document}


\section{Introduction}

The study of the Quark-Gluon Plasma (QGP) \cite{pasechnik2017} in heavy-ion collisions requires a precise knowledge of the initial state of the system in order to disentangle QGP-induced phenomena from other nuclear effects. In this regard, the weakly-interacting Z and W$^\pm$ bosons, when detected through their leptonic decay channels, provide a medium-blind reference that allows us to probe the initial-state effects such as the nuclear modifications of the Parton Distribution Functions (PDFs). The various sets of nuclear PDFs (nPDFs) available are currently suffering from large uncertainties due to the lack of experimental constraints in the $x$-range probed at the LHC \cite{paukkunen2011}.

Thanks to the high collision energies and luminosities delivered by the LHC, the measurement of the production of electroweak bosons is now accessible in heavy-ion collisions \cite{aliceZPbPb5tev, aliceWZpPb5tev, atlasZPbPb2tev, atlasZpPb5tev, lhcbZpPb5tev, cmsWPbPb2tev, cmsZPbPb2tev, cmsZpPb5tev}. The four main LHC experiments have complementary kinematic coverages that give access to a wide range of Bjorken-$x$ values (from $10^{-4}$ to almost unity) in a region of high virtuality ($Q^2 \sim M^2_{Z,W}$) where the nPDFs are poorly constrained by other experiments.

\section{Analysis and results}

\subsection{Procedure}

The Z and W$^\pm$ bosons are detected in their (di)muonic decay channels from data recorded with the ALICE muon spectrometer \cite{spectrometer}. The spectrometer is a conical-shape detector consisting of five stations of tracking chambers, two stations of trigger chambers and a set of absorbers that shield the detector from various background sources such as hadrons produced in the collision and the products of interactions of large-$\eta$ primary particles with the beam-pipe. A complete description of the ALICE detector can be found in \cite{alice}. The spectrometer acceptance covers the pseudo-rapidity interval $-4 < \eta < -2.5$, corresponding to a rapidity acceptance in the range $2.5 < y < 4$ for Pb--Pb collisions.\footnote{In the ALICE reference frame, the spectrometer covers negative $\eta$. However, positive values are used when referring to $y$.} In proton-lead collisions, the proton and lead beams have different energies, the nucleon-nucleon centre-of-mass is thus boosted with respect to the one in the laboratory frame by $\Delta y = 0.465$ in the direction of the proton beam. The rapidity acceptance of the spectrometer is then $2.03 < y_{cms} < 3.53$ ($-4.46 < y_{cms} < -2.96$) in the p--going (Pb--going) direction, when the proton (Pb) beam moves toward the spectrometer.

In the following, results from three different periods are presented. The collision systems, energies in the centre-of-mass, integrated luminosities and performed measurements are summarized in Table \ref{periods}.

\begin{table}[h]
  \centering
  \begin{tabular}{|c|c|c|c|c|}
    \hline
    \textbf{Collision system} & \textbf{Year}         & $\mathbf{\sqrt{s_{_{\rm \textbf{NN}}}}}$ & \textbf{Luminosity}                       & \textbf{Analyses} \\
    \hline \hline
    p--Pb	              & \multirow{2}{*}{2013} & \multirow{2}{*}{5.02 TeV}          & 5.03 $\pm$ 0.18 nb\textsuperscript{-1}    & \multirow{2}{*}{Z, W$^\pm$} \\
    Pb--p                     &                       &                                    & 5.81 $\pm$ 0.20 nb\textsuperscript{-1}    & \\
    \hline
    Pb--Pb                    & 2015                  & 5.02 TeV                           & $\sim$ 225 $\rm \mu$b\textsuperscript{-1} & Z \\
    \hline
    p--Pb	              & \multirow{2}{*}{2016} & \multirow{2}{*}{8.16 TeV}          & 8.47 $\pm$ 0.18 nb\textsuperscript{-1}    & \multirow{2}{*}{Z} \\
    Pb--p                     &                       &                                    & 12.75 $\pm$ 0.25 nb\textsuperscript{-1}   & \\
    \hline
  \end{tabular}
  \caption{Analyses performed in the different data taking periods.}
  \label{periods}
\end{table}

In order to ensure the quality of the data sample, a selection is applied on each muon track on the trigger-tracker matching (each track reconstructed in the tracking chambers has to match a track segment in the trigger chambers), the distance at the end of the front absorber and the product of the track momentum to its distance of closest approach (i.e. the distance to the primary vertex of the track extrapolated to the plane transverse to the beam axis and containing the vertex itself). This aims at reducing the background noise by removing tracks that are poorly reconstructed, crossing the high-Z material part of the detector or not coming from the interaction vertex.

The W$^\pm$ candidates are extracted by fitting the single muon transverse momentum distribution, starting at $\pt ^\mu > 10$ GeV/$c$ to remove muons from low-mass particles decay. The fit is performed using a combination of Monte-Carlo templates to account for the various contributions. In the Z case, a selection on the muon pseudo-rapidity ($-4 < \eta_\mu < -2.5$) and transverse momentum ($\pt ^\mu > 20$ GeV/$c$) as well as on the dimuon invariant mass ($60 < m_{\mu\mu} < 120$ GeV/$c^2$) leaves a nearly background-free sample from which the Z signal is obtained by simply counting the number of opposite-sign dimuons in the fiducial region defined by the selection.

Finally, the systematic uncertainties on the measurements are evaluated by combining the systematics on the signal extraction, including its potential contamination by background sources, the detector performances and the simulations from which the acceptance $\times$ efficiency factor is estimated.

The strategies for the Z and W$^\pm$ analyses are described in more details in \cite{aliceZPbPb5tev, aliceWZpPb5tev}.

\subsection{Pb--Pb collisions}

\begin{figure}
  \centering
  \includegraphics[height=7cm]{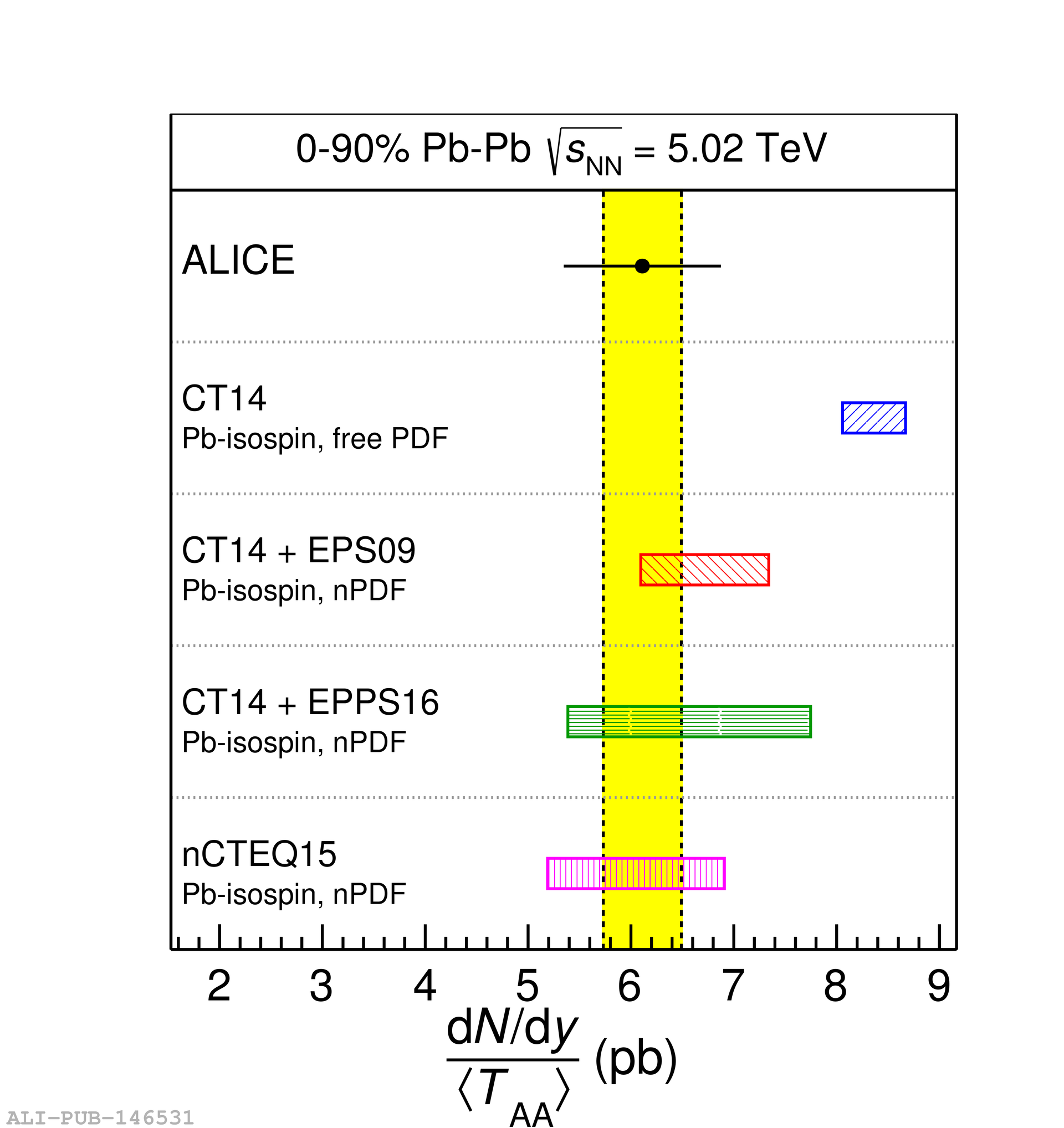}
  \includegraphics[height=6.4cm]{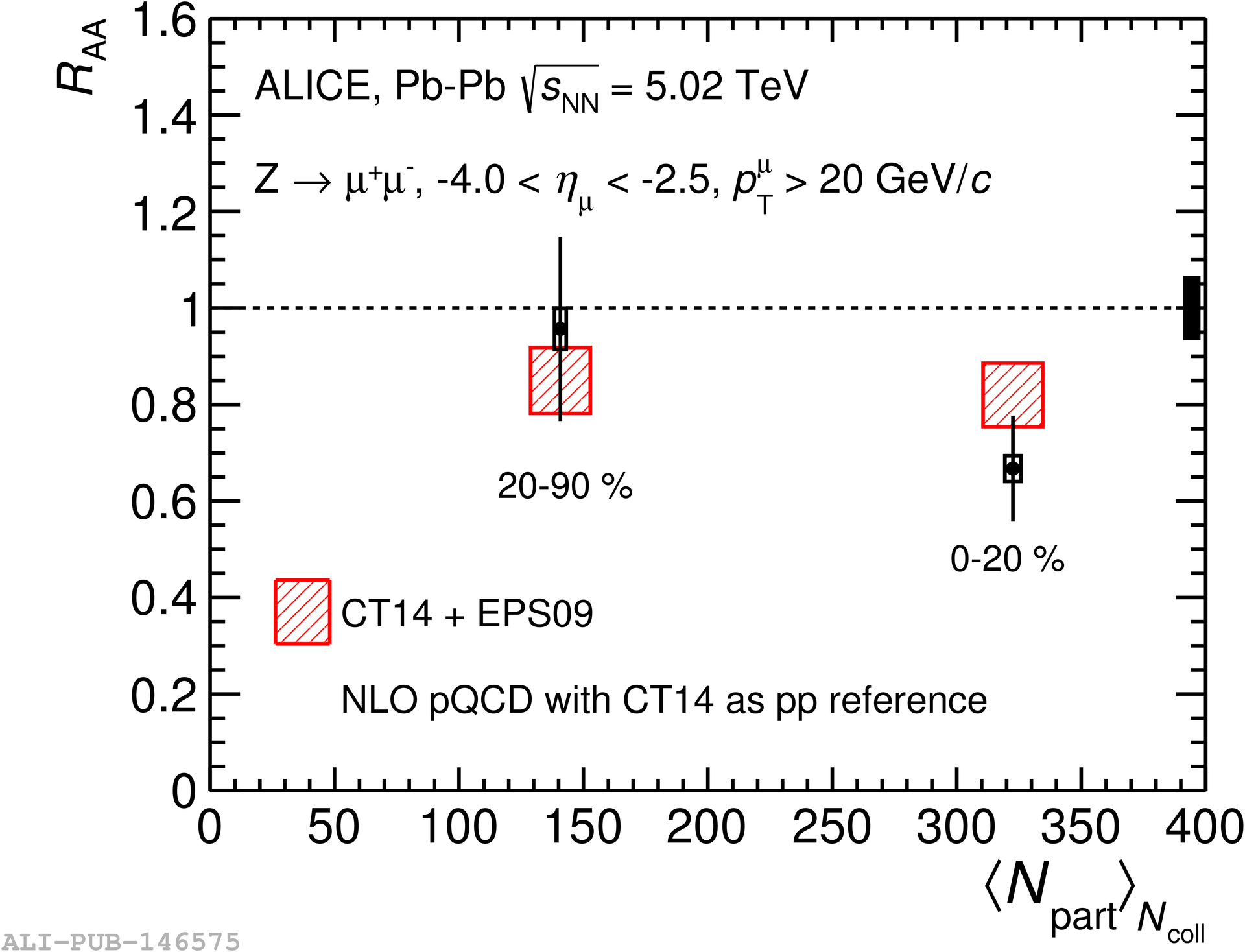}
  \caption{\textbf{Left}: Z yield in Pb--Pb collisions at \snn{} = 5.02 TeV, divided by the average nuclear overlap function, in the rapidity region $2.5 < y < 4$ and a $0-90\%$ collision centrality, compared to predictions from free-PDF and various nPDFs. The horizontal bar and boxes correspond to the statistical and model uncertainties, while the filled band indicates the systematic uncertainty of the experimental value. \textbf{Right}: Centrality dependence of the Z nuclear modification factor. The vertical bars represent the statistical uncertainty, while the boxes represent the systematic uncertainty on the measurements. The filled box located at $R_{AA} = 1$ shows the normalisation uncertainty. See text for details about the models.}
  \label{PbPbYield}
\end{figure}

The measured Z yield, normalized to the average nuclear overlap function $\left< T_{AA} \right>$ obtained from the Glauber model \cite{glauber}, is displayed in the left panel of Figure \ref{PbPbYield}, for a collision centrality in the $0-90\%$ range and integrated in rapidity ($2.5 < y < 4$). The measurement is compared to several predictions, from free-PDF only, using CT14 \cite{ct14}, or using three different parametrizations for the nuclear modification of the PDFs: EPS09 \cite{eps09}, EPPS16 \cite{epps16} (both with CT14 as baseline PDF) or nCTEQ15 \cite{ncteq15, ncteq15vecBos}. Calculations using the free-PDF set alone are found to overestimate the measured yield by $2.3\sigma$, while the predictions from nPDFs are all in agreement with the measurement within uncertainties.

Figure \ref{PbPbYield} (right) shows the centrality dependence of the Z nuclear modification factor $R_{AA}$, compared to calculations including a centrality-dependent nuclear modification of the PDFs. The nuclear modification factor is defined as the ratio of the yield in Pb--Pb collisions to the cross section in p--p collisions scaled by the average nuclear overlap function:
\begin{equation}
  R_{AA} = \frac{1}{\left< T_{AA} \right>} \frac{\dd N_{AA} / \dd y}{\dd \sigma _{pp} / \dd y}.
\end{equation}
A significant deviation by $3\sigma$ from unity is observed for the 20\% most central collisions. Theoretical calculations including the nuclear modifications of the PDFs are found to reproduce the measured ratio within uncertainties.

\subsection{p--Pb collisions}

\begin{figure}
  \centering
  \includegraphics[width=0.45\linewidth]{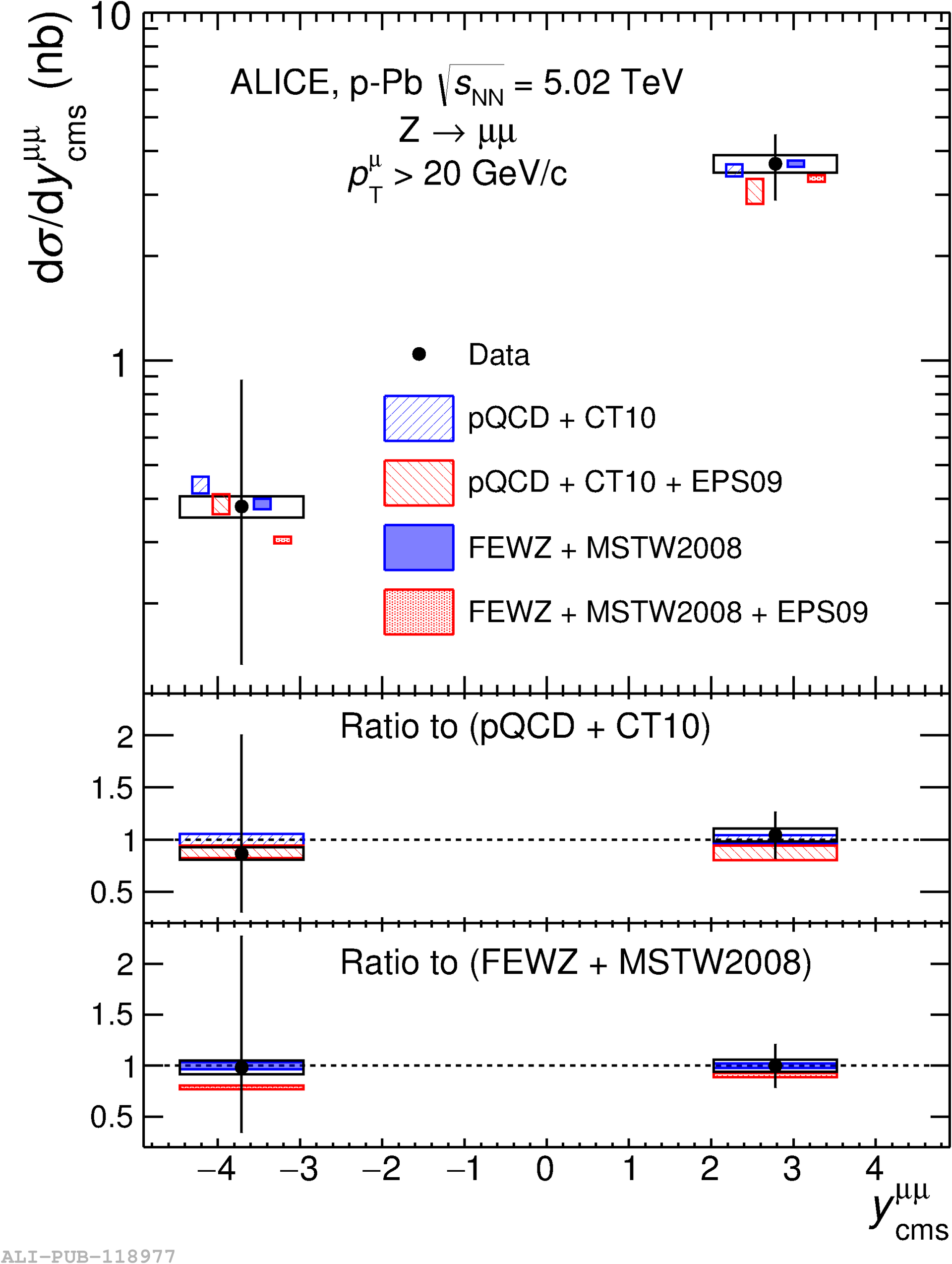}
  \includegraphics[width=0.45\linewidth]{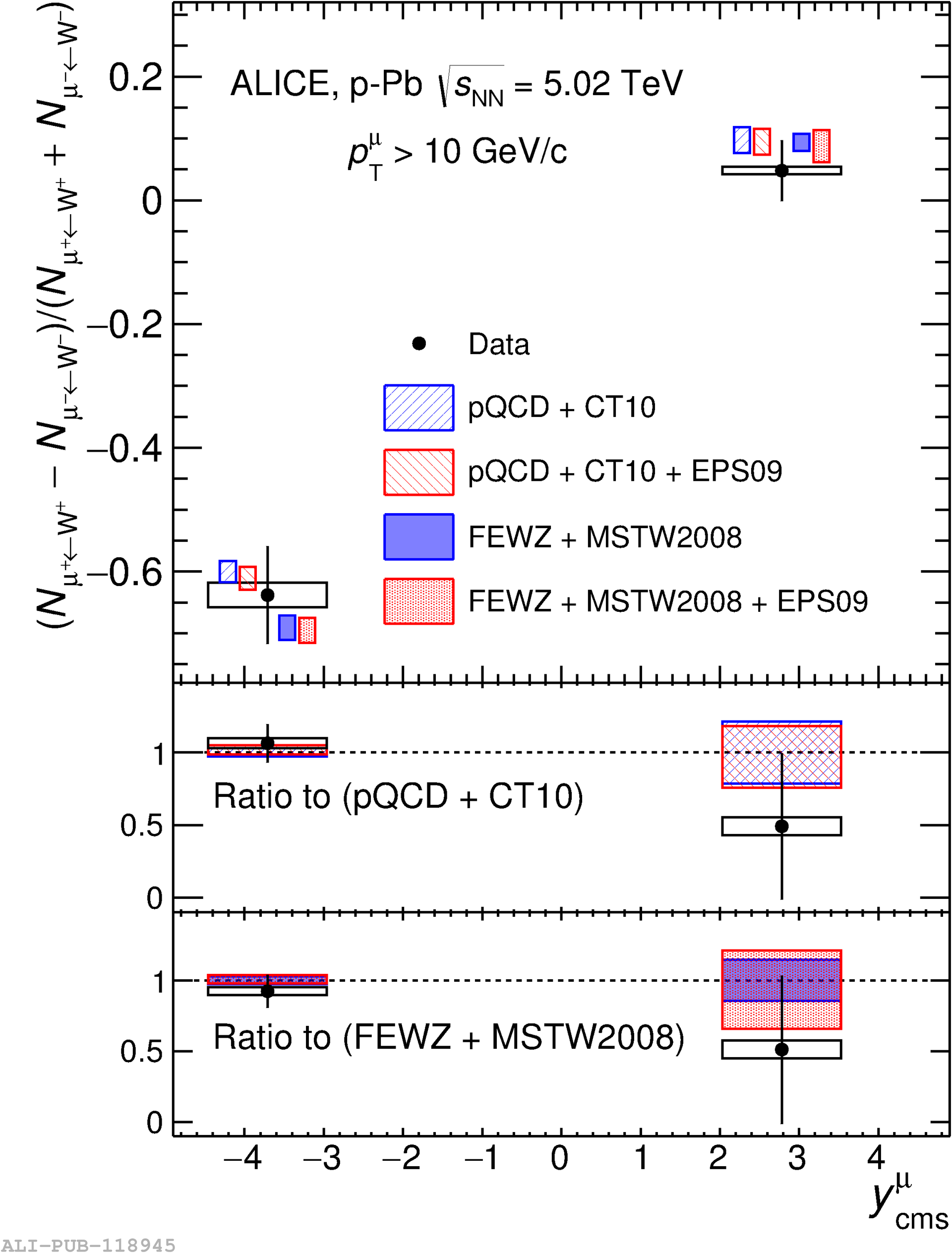}
  \caption{\textbf{Left}: Z production as a function of rapidity in p--Pb collisions at \snn{} = 5.02 TeV. The bars and boxes around the data points correspond to statistical and systematic uncertainties respectively. The horizontal width of the boxes indicates the measured rapidity range. The measurements are compared to theoretical predictions, horizontally shifted for readability. \textbf{Right}: W lepton charge asymmetry in p--Pb collisions at \snn{} = 5.02 TeV compared to theoretical calculations, at forward and backward rapidities. The statistical (systematic) uncertainties are displayed as bars (boxes), the predictions are shifted horizontally for readability. In both figures, the vertical middle (bottom) panels display the ratio of the data and NLO (NNLO) calculations with nuclear modifications over the NLO (NNLO) calculations from free-PDF only.}
  \label{pPb5tev}
\end{figure}

Figure \ref{pPb5tev} (left) shows the Z production cross sections measured in p--Pb and Pb--p collisions at \snn{} = 5.02 TeV. They are compared to theoretical calculations at NLO and NNLO, both with and without nuclear modifications of the PDF. The NLO calculations are obtained from pQCD, using CT10 \cite{ct10} as free-PDF. The NNLO predictions are computed in the FEWZ \cite{fewz} framework with MSTW2008 \cite{mstw2008} as baseline PDF. In both cases, the nuclear modifications are implemented with EPS09 \cite{eps09}. The ratios of the measured cross sections to the predictions with nuclear modification are shown in the middle (bottom) panel of the figure for NLO (NNLO) computations.

Figure \ref{pPb5tev} (right) presents the W lepton-charge asymmetry, a quantity that gives access to the down over up quark ratio while allowing for a partial cancellation of the uncertainties. It is compared to predictions obtained from the same methods as for the Z cross section computation. The measurement-to-theory ratio can be seen in the lowest panels of each figure.

The effect of the nuclear modifications is smaller in p--Pb collisions than in the Pb--Pb case. In addition, the measured yields are found to be smaller, leading to a higher associated relative uncertainty. The combination of those effects prevents any firm conclusion on nuclear effects, as the measured cross sections and charge asymmetry are found to be consistent, within uncertainties, with predictions both including and excluding nuclear modifications.\\

\begin{figure}
  \centering
  \includegraphics[width=0.42\linewidth]{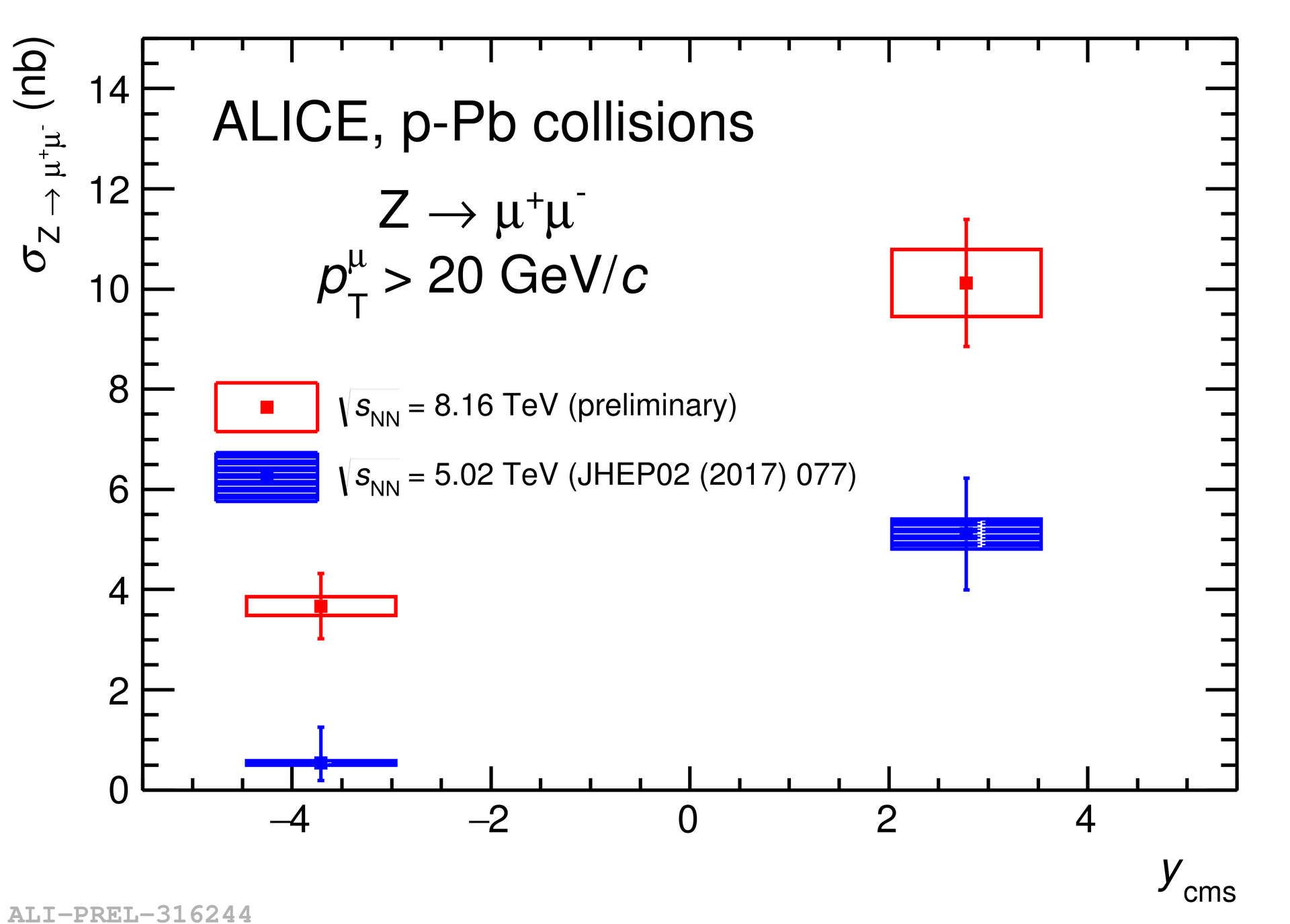}
  \includegraphics[width=0.45\linewidth]{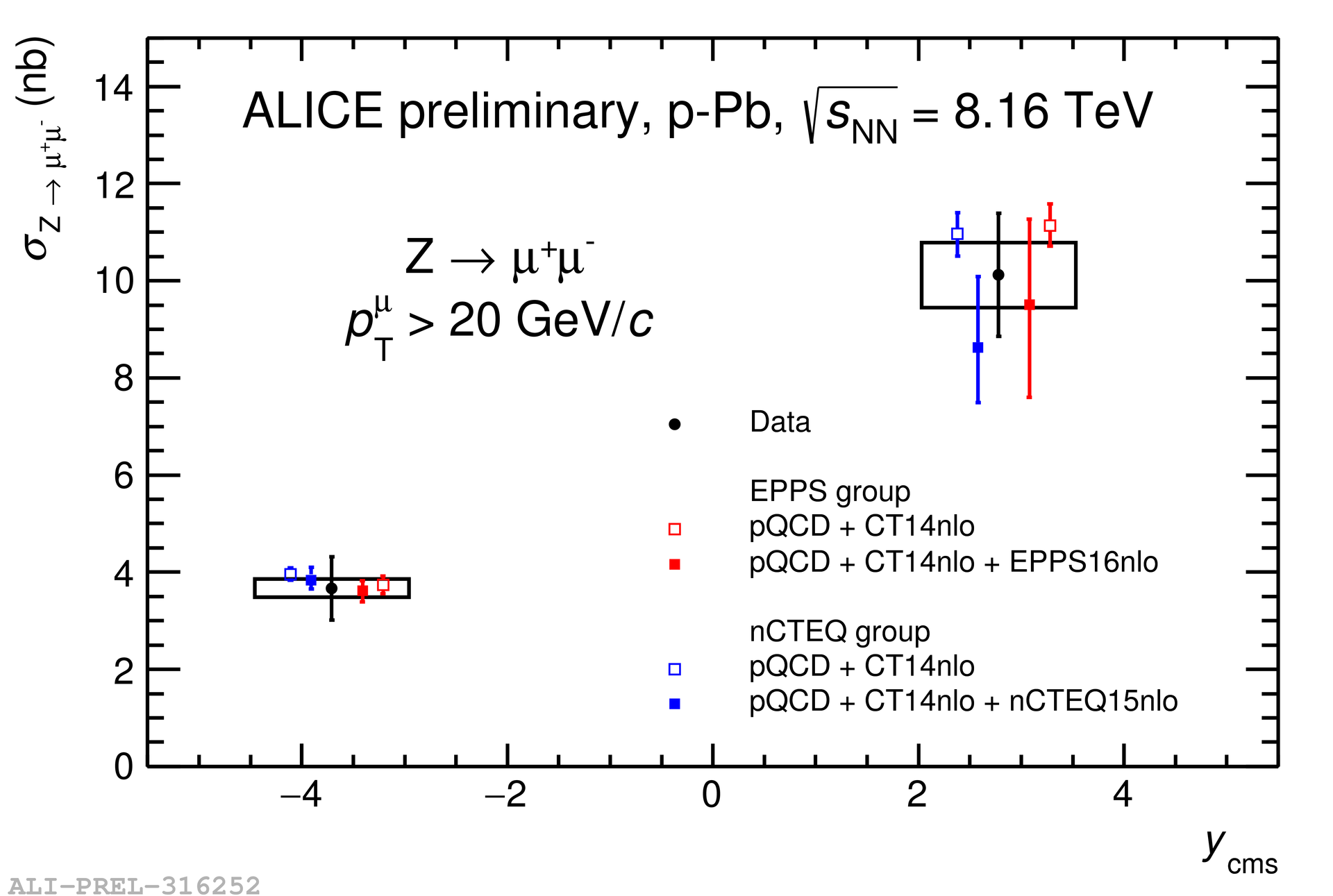}
  \caption{Z production cross section as a function of rapidity in p--Pb collisions at \snn{} = 8.16 TeV, compared to the measured production at 5.02 TeV (left panel) and theoretical calculations (right panel). The bars and boxes around the data points correspond to statistical and systematic uncertainties respectively. The theoretical points are horizontally shifted for readability, the close (open) symbols correspond to predictions with (without) nuclear modification of the PDF.}
  \label{pPb8tev}
\end{figure}

A new preliminary cross section of the Z production has been measured in p--Pb and Pb--p collisions at \snn{} = 8.16 TeV. The results are displayed in Figure \ref{pPb8tev} (left) where they are compared to the 5.02 TeV measurement. The same backward -- forward asymmetry is observed, which results from the change in Bjorken-$x$ induced by the rapidity shift. Figure \ref{pPb8tev} (right) shows the measured cross sections, compared with theoretical calculations at NLO, with and without including nuclear effects. The predictions with nuclear modifications are calculated using the two most recents nPDFs available, EPPS16 \cite{epps16}, using CT14 \cite{ct14} as baseline PDF, and nCTEQ15 \cite{ncteq15}. Within uncertainties, the measurements are reproduced with and without nuclear modifications of the PDFs.

\section{Conclusion}

The electroweak-boson measurements performed by the ALICE collaboration have been presented. In Pb--Pb at \snn{} = 5.02 TeV, the measured Z production is well reproduced by theoretical calculations including nuclear modifications of the PDFs, while a significant deviation is found from free-PDF predictions, by $2.3\sigma$ for the centrality-integrated yield. In p--Pb collisions, at centre-of-mass energies of 5.02 and 8.16 TeV, the measurements are well reproduced by calculations but the statistical limitations an the small magnitude of nuclear effects prevent any firm conclusion to be drawn on nuclear modifications. Further measurements with better precision are needed to provide more stringent constraints on the nPDFs.

\bibliographystyle{JHEP}
\bibliography{proceeding_eps-hep-2019}

\end{document}